\documentclass[aip,jcp,twocolumn,superscriptaddress]{revtex4}%
\usepackage{epsfig,dsfont,amssymb,amsmath,amsthm,amsfonts,amsbsy,mathrsfs}
\usepackage{graphicx}
\usepackage{amsmath}
\usepackage{amssymb}%
\begin{document}

\title{Defect in Phosphorene}

\author{Wei Hu}
\thanks{Corresponding author. E-mail: whu@lbl.gov}
\affiliation{Computational Research Division, Lawrence Berkeley
National Laboratory, Berkeley, CA 94720, USA}




\author{Jinlong Yang}
\thanks{Corresponding author. E-mail: jlyang@ustc.edu.cn}
\affiliation{Hefei National Laboratory for Physical Sciences at
Microscale, University of Science and Technology of China, Hefei,
Anhui 230026, China} \affiliation{Synergetic Innovation Center of
Quantum Information and Quantum Physics, University of Science and
Technology of China, Hefei, Anhui 230026, China}

\date{\today}

\pacs{ }

\begin{abstract}

Defects are inevitably present in materials and always can affect
their properties. Here, first-principles calculations are performed
to systematically study the stability, structural and electronic
properties of ten kinds of point defects in semiconducting
phosphorene, including the Stone-Wales (SW-1 and SW-2) defect,
single (SV59 and SV5566) and double vacancy (DV585-1, DV585-2,
DV555777-1, DV555777-2, DV555777-3 and DV4104) defects. We find that
these defects are all much easily created in phosphorene with higher
areal density compared with graphene and silicene. They are easy
distinguish each other and correlate with their defective atomic
structures with simulated scanning tunneling microscopy images at
positive bias. The SW, DV585-1, DV555777 and DV4104 defects have
little effect on phosphorene's electronic properties and defective
phosphorene monolayers still show semiconducting with similar band
gap values (about 0.9 $eV$) to perfect phosphorene. The SV59 and
DV585-2 defects can introduce unoccupied localized states into
phosphorene's fundamental band gap. Specifically, the SV59 and 5566
defects can induce hole doping in phosphorene, and only the stable
SV59 defect can result in local magnetic moments in phosphorene
different from all other defects.

\end{abstract}

\maketitle

\section{Introduction}

Two-dimensional (2D) ultrathin materials, such as
graphene,\cite{NatureMater_6_183_2007}
silicene,\cite{PRL_102_236804_2009} hexagonal boron
nitride,\cite{NatureMater_3_404_2004} graphitic carbon
nitride\cite{NatureMater_8_76_2009} and molybdenum
disulphide,\cite{NatureNanotechnol_6_147_2011} have received
considerable interest recently owing to their outstanding properties
and potential applications. Graphene,\cite{NatureMater_6_183_2007} a
2D sp$^2$-hybridized carbon monolayer, is known to have remarkable
electronic properties, such as a high carrier mobility, but the
absence of a bandgap limits its applications of large-off current
and high on-off ratio for graphene-based electronic devices. The
same problem also exists in silicene, another well-known single
silicon monolayer, which has most similar remarkable electronic
properties to graphene but with buckled honeycomb
structures.\cite{PRL_102_236804_2009}

Most recently, a new 2D material, namely,
phosphorene,\cite{PRB_86_035105_2012, JPCC_118_14051_2014,
NatureNanotech_9_372_2014, NatureCommun_5_4475_2014,
ACSNano_8_4033_2014} has been isolated in the experiments through
mechanical exfoliation from bulk black phosphorus and has
immediately received considerable attention. Phosphorene also shows
some remarkable electronic properties superior to graphene and
silicene. For example, phosphorene is a semiconductor with a direct
bandgap of about 1 $eV$,\cite{JPCC_118_14051_2014} showing the drain
current modulation up to 10$^5$ and carrier mobility up to 10$^3$
cm$^2$/(Vs) in nanoelectronics.\cite{NatureNanotech_9_372_2014,
NatureCommun_5_4475_2014, ACSNano_8_4033_2014} Furthermore, up to
now, besides graphene, phosphorene is the only stable elemental 2D
material which can be mechanically exfoliated in the
experiments.\cite{ACSNano_8_4033_2014} Therefore, introduced as an
alternative to graphene,\cite{ScienceNews_185_13_2014} phosphorene
may lead to faster semiconductor electronics in the future.

On the other hand, the properties of these 2D ultrathin materials
are always affected by the presence of defects.\cite{ACR_2014}
Typical point defects in graphene and silicene include Stone-Wales
(SW) defect, single and double vacancy (SV and DV)
defects.\cite{Nature_430_870_2004, NanoLett_7_2459_2007,
ACSNano_5_26_2011, Nanoscale_5_9785_2013, PRB_88_045434_2013}
Generally, both graphene and silicene have two kinds of SVs (SV59
and SV5566) and DVs (DV585 and DV555777), respectively. These
defects are inevitably present in graphene and silicene and severely
affect their structural and electronic
properties,\cite{ACSNano_5_26_2011, Nanoscale_5_9785_2013} thus
alter their applications.\cite{APL_93_193107_2008,
Nanoscale_5_9062_2013, PCCP_15_5753_2013} However, up to now, little
attention has been focused on the defects in
phosphorene.\cite{NanoLett_14_6782_2014}

In the present work, we systematically study the stability and
electronic structures of typical point defects in semiconducting
phosphorene using the density functional theory and ab-initio
molecular dynamics calculations. We find that phosphorene has a wide
variety of defects due to its low symmetry structure. Furthermore,
these defects are all much easily created in phosphorene with regard
to graphene and silicene and they are easy distinguish each other
and correlate with their defective atomic structures with simulated
scanning tunneling microscopy images at positive bias. Defects of
different structures shows different stability and electronic
structures in in phosphorene.

\section{Theoretical Methods and Models}

First-principles calculations are based on the density functional
theory (DFT) implemented in the VASP
package.\cite{PRB_47_558_1993_VASP} The generalized gradient
approximation of Perdew, Burke, and Ernzerhof
(GGA-PBE)\cite{PRL_77_3865_1996_PBE} is chosen due to its good
description of electronic structures of
phosphorene,\cite{JPCC_118_14051_2014}
graphene\cite{ACSNano_5_26_2011} and
silicene.\cite{Nanoscale_5_9785_2013} All the geometry structures
are fully relaxed until energy and forces are converged to 10$^{-5}$
$eV$ and 0.01 $eV$/{\AA}, respectively. The energy cutoff is set to
be 500 $eV$. The lattice parameters of phosphorene calculated to
setup unit cell are $a$(P) = 4.62 {\AA} and $b$(P) = 3.30
{\AA}.\cite{JPCC_118_14051_2014} A large 5 $\times$ 7 supercell of
phosphorene with a vacuum space about 15 {\AA} in the Z direction is
adopted to study the effect of various local defects in phosphorene.
The surface Brillouin zone is sampled with a 3 $\times$ 3 regular
mesh.

Ab-initio molecular dynamics (AIMD) simulations are employed to
study the stability of defects in phosphorene. AIMD simulations are
performed in a canonical ensemble. The simulations is performed
during 2.0 $ps$ with a time step of 2.0 $fs$ at the temperature of
400 $K$ controlled by using the Nose-Hoover
method.\cite{PRB_51_13705_1995_NoseHoover}

\section{Results and Discussion}

\begin{figure}[htbp]
\begin{center}
\includegraphics[width=0.5\textwidth]{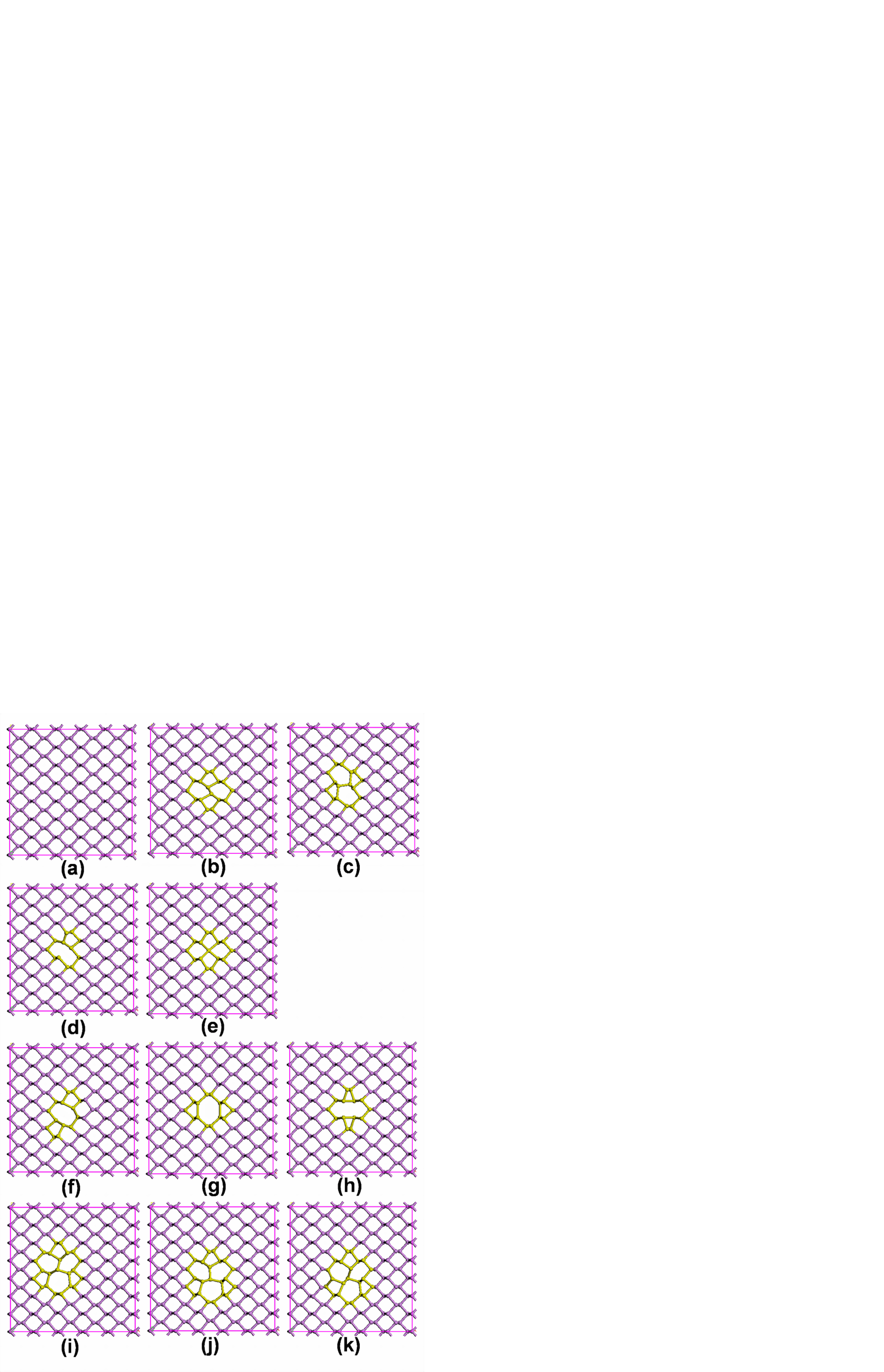}
\end{center}
\caption{(Color online) Geometric structures of (a) perfect and
defective phosphorene in the 5 $\times$ 7 supercell, including the
(b) SW-1, (c) SW-2, (d) SV59, (e) SV5566, (f) DV585-1, (g) DV585-2,
(h) DV4104, (i) DV555777-1, (j) DV555777-2 and (k) DV555777-3
defects. The violet and yellow balls denote unaffected and affected
phosphorus atoms, respectively.} \label{fig:Structure}
\end{figure}

We first check the geometric properties of defects in phosphorene as
shown in Figure~\ref{fig:Structure}. Typical point defects of SW,
SV59, SV5566, DV585 and DV555777 in graphene\cite{ACSNano_5_26_2011}
and silicene\cite{Nanoscale_5_9785_2013} also can be formed in
phosphorene, but which has a wide variety of defects because of
lower symmetry structure. There are two kinds of SW (SW-1 and SW-2)
and DV585 (DV585-1 and DV585-2) defects respectively, three kinds of
DV555777 (DV555777-1, DV555777-2 and DV555777-3) defects in
phosphorene as shown in Figure~\ref{fig:Structure}. Interestingly,
we find a new stable structure of DV defect in phosphorene, namely
4$|$10$|$4, as shown in Figure~\ref{fig:Structure}(h), which has not
been observed in both graphene and silicene.

\begin{figure}[htbp]
\begin{center}
\includegraphics[width=0.5\textwidth]{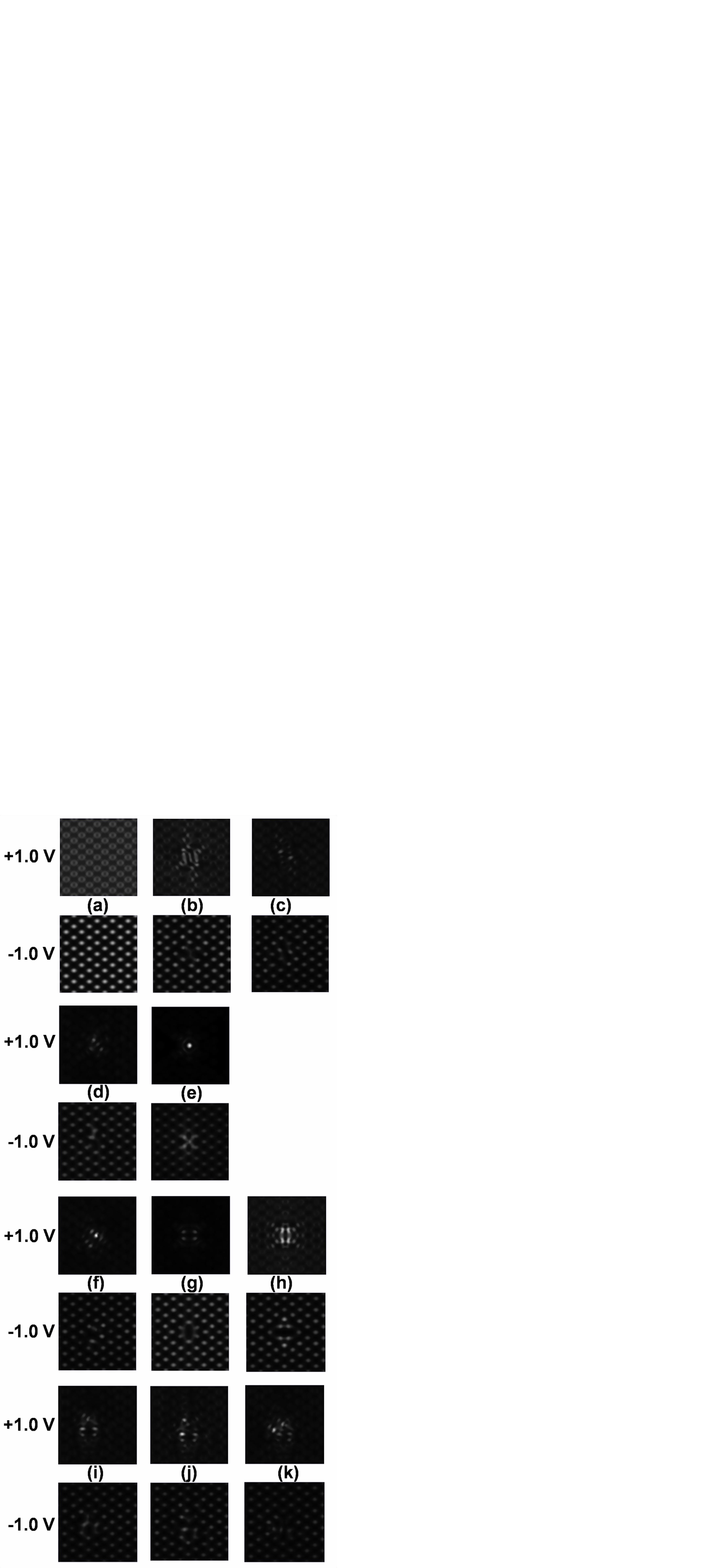}
\end{center}
\caption{(Color online) Simulated STM images (+1.0 and -1.0 $V$) of
(a) perfect and defective phosphorene with the (b) SW-1, (c) SW-2,
(d) SV59, (e) SV5566, (f) DV585-1, (g) DV585-2, (h) DV4104, (i)
DV555777-1, (j) DV555777-2 and (k) DV555777-3
defects.}\label{fig:STM}
\end{figure}

To help recognize the defects in future experiments, the scanning
tunneling microscopy (STM) images of perfect and defective
phosphorene are simulated at +1.0 and -1.0 $V$ bias as shown in
Figure~\ref{fig:STM}. At positive bias (+1.0 $V$), the STM images of
these defects are easy to understand and correlate with their
defective atomic structures. But at negative bias (-1.0 $V$), the
STM images of perfect and defective phosphorene are difficult to
distinguish each other due to a direct consequence of height
variation for the buckling of phosphorene.

Then, we study the stability of defects in phosphorene. To
characterize the stability of a defect in phosphorene, we define its
formation energy as
\[
E_{f}=E_{\text{\tiny Phosphorene}}-N_P*E_{P}
\]
where $E_{\text{\tiny Phosphorene}}$ represents the total energy of
defective phosphorene, $E_{P}$ is the energy per phosphorus atom in
a perfect phosphorene sheet and $N_P$ corresponds to the number of
phosphorus atoms in phosphorene. Notice that for perfect
phosphorene, $E_{f}$(Phosphorene) = 0 $eV$. Perfect phosphorene
shows less stable with a smaller cohesive energy of 3.48 $eV$/atom
compared with graphene (7.90 $eV$/atom) and silicene (3.96
$eV$/atom).\cite{Nanoscale_5_9785_2013} Our calculated stability of
perfect and defective phosphorene, graphene and silicene are
summarized in Table~\ref{table1} and Table~\ref{table2}.

\begin{table}
\caption{Calculated stability and electronic properties of perfect
and defective phosphorene, including the formation energy $E_f$
($eV$), total magnetic moment $\mu$ ($\mu$$_B$) and band gap $E_g$
($eV$).} \label{table1}
\begin{tabular}{cccc} \\ \hline \hline
              & $E_f$  &  $\mu$  &  $E_g$      \ \\
\hline
Perfect       & 0.000   &  0.000  &  0.905       \ \\
SW-1          & 1.012   &  0.003  &  0.928       \ \\
SW-2          & 1.322   &  0.003  &  0.883       \ \\
SV59          & 1.626   &  0.980  &  0.190/0.941 \ \\
SV5566        & 2.025   &  0.000  &  Hole doping \ \\
DV585-1       & 1.906   &  0.002  &  0.962       \ \\
DV585-2       & 3.041   &  0.002  &  0.559       \ \\
DV4104        & 2.137   &  0.000  &  0.929       \ \\
DV555777-1    & 2.081   &  0.002  &  0.966       \ \\
DV555777-2    & 2.350   &  0.002  &  0.962       \ \\
DV555777-3    & 2.613   &  0.003  &  0.973       \ \\
\hline \hline
\end{tabular}
\end{table}

\begin{table}
\caption{Comparison results of cohesive energy $E_c$ ($eV$/atom) of
perfect phosphorene, graphene and silicene with corresponding
various defects' formation energy $E_f$ ($eV$).} \label{table2}
\begin{tabular}{cccc} \\ \hline \hline
                 &  Phosphorene  &  Graphene  &  Silicene   \ \\
Reference        &  This work    &  Ref.\cite{ACSNano_5_26_2011}  &  Ref.\cite{Nanoscale_5_9785_2013}   \ \\
\hline
$E_c$(Perfect)   &  3.48         &  7.90      &   3.96      \ \\
$E_f$(SW)        &  1.01$-$1.32  &  4.50      &   2.09      \ \\
$E_f$(SV59)      &  1.63         &  7.80      &   3.77      \ \\
$E_f$(SV5566)    &  2.03         &  $-$       &   3.01      \ \\
$E_f$(DV585)     &  1.91$-$3.04  &  7.52      &   3.70      \ \\
$E_f$(DV555777)  &  2.08$-$2.61  &  6.40      &   2.84      \ \\
$E_f$(DV4104)    &  2.13         &  $-$       &   $-$       \ \\

\hline \hline
\end{tabular}
\end{table}

We find that the SW-1 defect is most easily formed in phosphorene
with the smallest formation energy of 1.01 $eV$ among various
defects similar to graphene\cite{ACSNano_5_26_2011} and
silicene.\cite{Nanoscale_5_9785_2013}  For DVs, the 585 and 555777
defects are also stable in phosphorene similar to graphene and
silicene. Interestingly, we find the most stable DV in phosphorene
is 4104 with a small formation energy of 2.13 $eV$, but which can
not formed in graphene and silicene. Notice that defective
phosphorene monolayers all have smaller formation energy compared
with defective graphene and silicene with the same types of defects
as summarized in Table~\ref{table2}.

To further study the thermal stability of perfect and defective
phosphorene, AIMD simulations are performed at 400 $K$. In the
initial state ($t$ = 0.0 $ps$), perfect and defective phosphorene
monolayers are set to optimized geometric structures. We find that
the perfect phosphorene and most of phosphorene monolayers are
stable at 400 $K$ during $t$ = 2.0 $ps$. The only unstable one is
the SV5566 defect in phosphorene, which trends to change into more
stable SV59 defect in phosphorene at 400 $K$.

\begin{figure}[htbp]
\begin{center}
\includegraphics[width=0.5\textwidth]{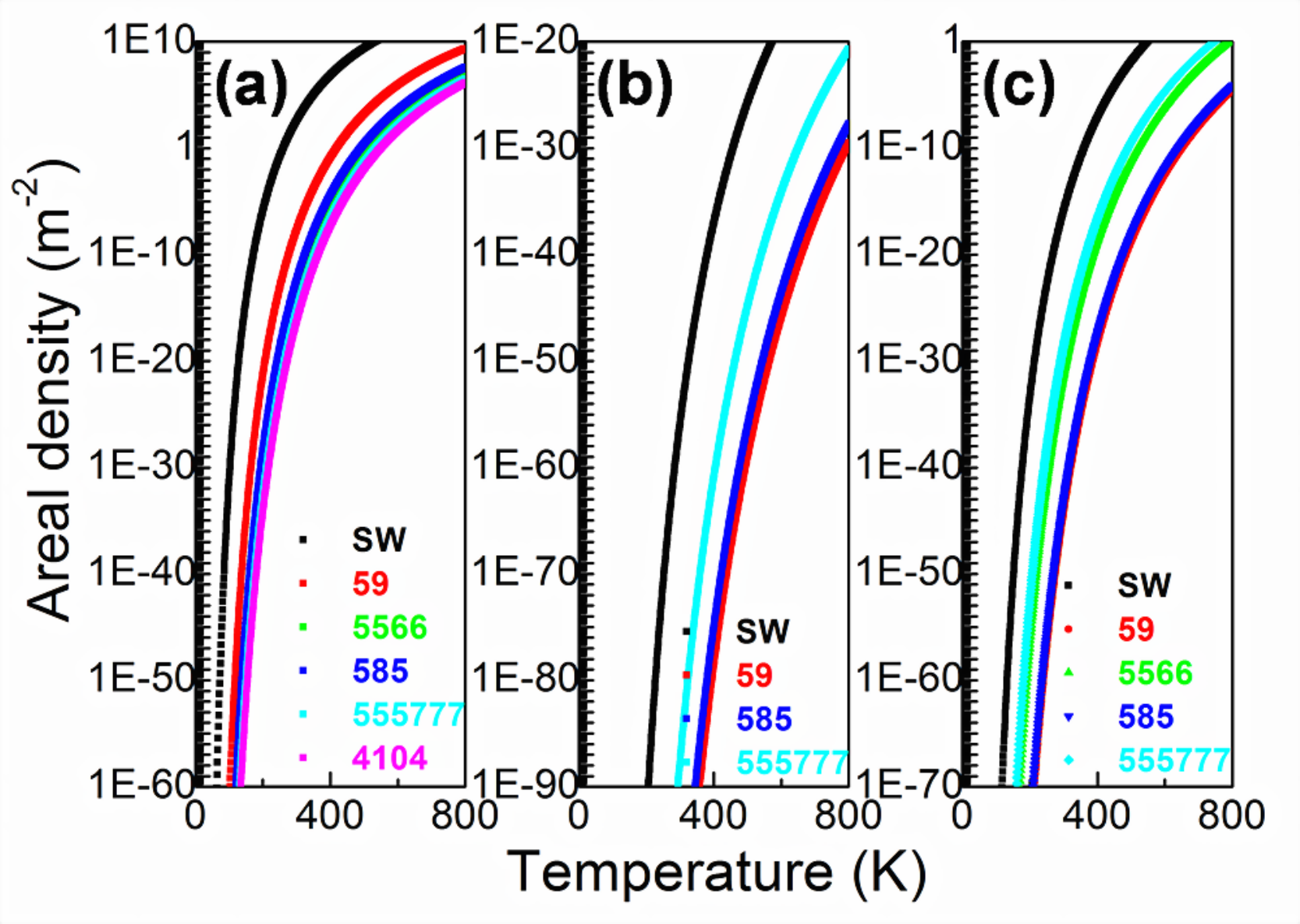}
\end{center}
\caption{(Color online) Areal density of various stable defects (SW,
SV59, SV5566, DV585, DV555777 and DV4104) in (a) phosphorene, (b)
graphene and (c) silicene as a function of
temperature.}\label{fig:Density}
\end{figure}

At finite temperature $T$, defects' areal density $N_{\text{\tiny
Defect}}$ ($m^{-2}$) in 2D materials follows the Arrhenius equation
\[
N_{\text{\tiny Defect}}=N_{\text{\tiny Perfect}}\exp(-E_f/k_{B}T)
\]
where $N_{\text{\tiny Perfect}}$ is the areal density of atoms in
perfect 2D materials, $E_f$ is the formation energy of a defect
formed in materials and $k_{B}$ is the Boltzmann constant. For
perfect phosphorene, graphene and silicene, their areal densities
are $N_{\text{\tiny Perfect}}$(Phosphorene) = 2.62 $\times$
$10^{19}$ $m^{-2}$, $N_{\text{\tiny Perfect}}$(Graphene) = 3.79
$\times$ $10^{19}$ $m^{-2}$ and $N_{\text{\tiny Perfect}}$(Silicene)
= 1.55 $\times$ $10^{19}$ $m^{-2}$, respectively. The
temperature-dependence areal density of the most stable defects in
phosphorene, graphene and silicene is calculated and plotted in
Figure~\ref{fig:Density}. The results show that these defects have
much higher areal density and they are much easily created in
phosphorene compared with graphene and silicene.

\begin{figure}[htbp]
\begin{center}
\includegraphics[width=0.5\textwidth]{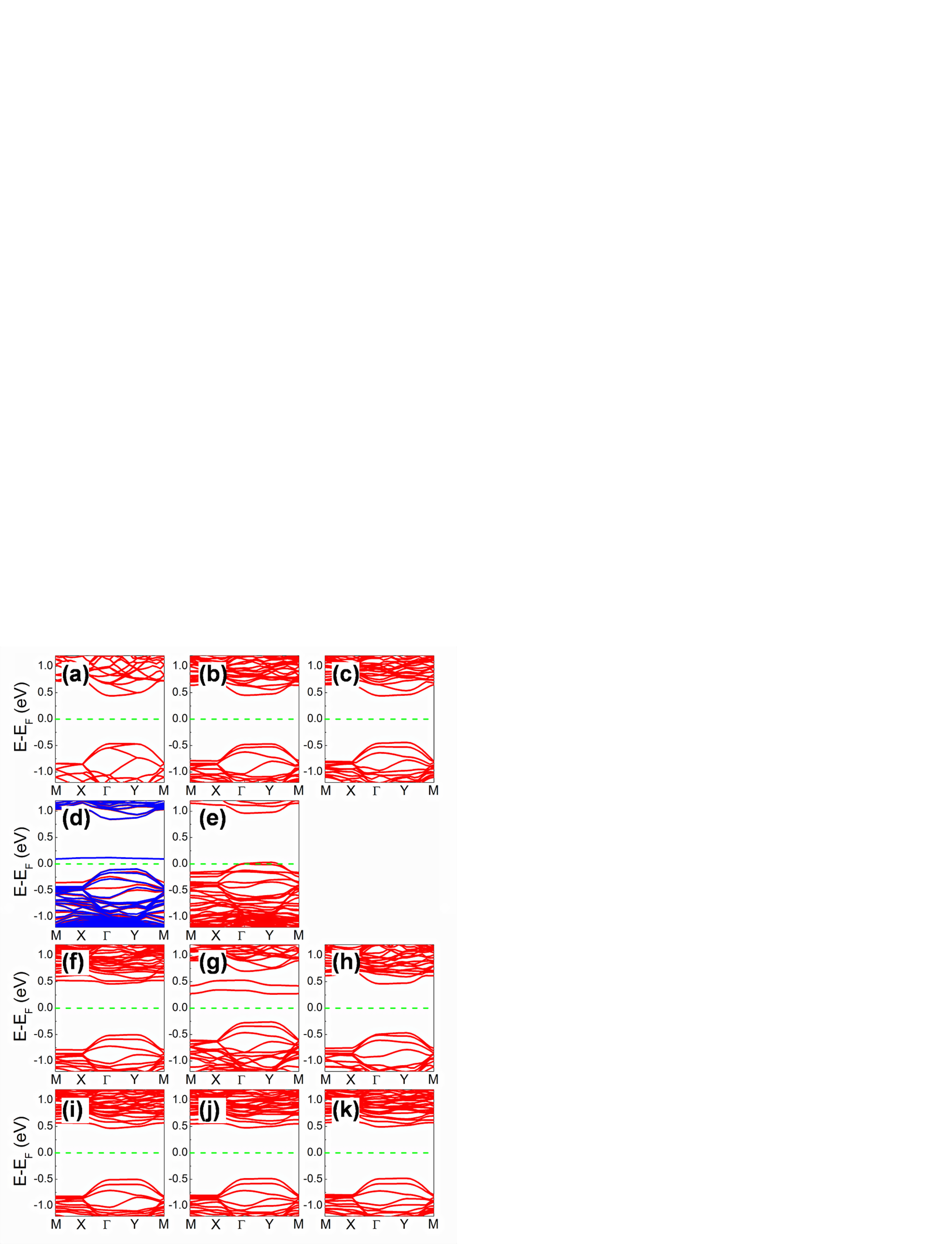}
\end{center}
\caption{(Color online) Electronic band structures of (a) perfect
and defective phosphorene with the (b) SW-1, (c) SW-2, (d) SV59, (e)
SV5566, (f) DV585-1, (g) DV585-2, (h) DV4104, (i) DV555777-1, (j)
DV555777-2 and (k) DV555777-3 defects. The Fermi level is marked by
green dotted lines and set to zero.}\label{fig:Band}
\end{figure}

Finally, we check the electronic band structures of perfect and
defective phosphorene as shown in Figure~\ref{fig:Band}. Monolayer
phosphorene is semiconducting with a direct band gap of 0.91 $eV$
(Figure~\ref{fig:Band}(a)), which agrees well with previous
theoretical studies.\cite{JPCC_118_14051_2014} We find that the SV59
defect can induce local magnetic moments in phosphorene with a
magnetic moment of 0.98 $\mu$$_B$, but other defective phosphorene
monolayers are not magnetic similar to
silicene.\cite{Nanoscale_5_9785_2013} Furthermore, the SW, DV555777
and DV4104 defects have little effect on phosphorene's electronic
properties, still showing semiconducting with similar band gap
values (about 0.9 $eV$) for defective phosphorene, different from
graphene\cite{ACSNano_5_26_2011} and
silicene.\cite{Nanoscale_5_9785_2013} Defective phosphorene with the
DV585 defects are also semiconductors but with different band gaps
(0.96 and 0.56 $eV$ respectively for DV585-1 and DV585-2). The SV59
and DV585-2 defects can introduce unoccupied localized states into
phosphorene's fundamental band gap as shown in
Figure~\ref{fig:Band}(d) and (g). Specifically, the SV defects can
induce hole doping in phosphorene as shown in
Figure~\ref{fig:Band}(d) and (e), which can increase the hole
carrier concentration of semiconducting phosphorene. Notice that
phosphorene has a high areal density for the SV defects, agreeing
well with a high hole mobility observed in phosphorene in recent
experiments.\cite{ACSNano_8_4033_2014}

\section{Conclusions}

In summary, we systematically study the stability and electronic
structures of defects in semiconducting phosphorene using the
density functional theory and ab-initio molecular dynamics
calculations. We find that phosphorene has a wide variety of point
defects (SW-1, SW-2, SV59, SV5566, DV585-1, DV585-2, DV555777-1,
DV555777-2, DV555777-3 and DV4104) due to its low symmetry
structure. Furthermore, these defects are all much easily created in
phosphorene with regard to graphene and silicene and they are easy
distinguish each other and correlate with their defective atomic
structures with simulated scanning tunneling microscopy images at
positive bias. Defects of different structures shows different
stability and electronic structures in in phosphorene. The SW,
DV585-1, DV555777 and DV4104 defects have little effect on
phosphorene's electronic properties and defective phosphorene
monolayers still show semiconducting with similar band gap values
(about 0.9 $eV$) to perfect phosphorene. The SV59 and DV585-2
defects can introduce unoccupied localized states into phosphorene's
fundamental band gap. Specifically, the SV59 and 5566 defects can
induce hole doping in phosphorene, and only the stable SV59 defect
can result in local magnetic moments in phosphorene different from
all other defects. The present theoretical results provide valuable
insights into the identification of defects in further experiments
and the understanding their effects on the properties and
applications of phosphorene.

\section{ACKNOWLEDGMENTS}

This work is partially supported by the National Key Basic Research
Program (2011CB921404), by NSFC (11404109, 21121003, 91021004,
21233007, 21222304), by CAS (XDB01020300). This work is also
partially supported by the Scientific Discovery through Advanced
Computing (SciDAC) program funded by U.S. Department of Energy,
Office of Science, Advanced Scientific Computing Research and Basic
Energy Sciences (W. H.). We thank the National Energy Research
Scientific Computing (NERSC) center, USTCSCC, SC-CAS, Tianjin, and
Shanghai Supercomputer Centers for the computational resources.

\end{document}